\documentclass[journal=jacsat,manuscript=article,layout=twocolumn]{achemso}

\SectionsOn
\SectionNumbersOn

\usepackage[version=3]{mhchem} % Formula subscripts using \ce{}
\usepackage[]{units}
\usepackage{graphicx}
\usepackage{xcolor}

\usepackage{color}
\usepackage{hyperref, soul}
\definecolor{linkColor}{rgb}{0,0,1}
\definecolor{citeColor}{rgb}{0,0,1}
\hypersetup{pdfborder={0 0 0}, colorlinks=true,urlcolor=linkColor,citecolor=citeColor}

\usepackage{cuted}

\usepackage{lineno}

\author{Frances I. Allen}
\affiliation{Department of Materials Science and Engineering, University of California, Berkeley, CA 97420, USA}
\alsoaffiliation{California Institute for Quantitative Biosciences, University of California, Berkeley, CA 97420, USA}
\alsoaffiliation{National Center for Electron Microscopy, Molecular Foundry, Lawrence Berkeley National Laboratory, Berkeley, CA 97420, USA}
\email{francesallen@berkeley.edu}

\author{Jos\'e Mar\'ia De Teresa}
\affiliation{Instituto de Nanociencia y Materiales de Aragón (INMA), CSIC-Universidad de Zaragoza, 50009, Spain}

\author{Bibiana Onoa}
\affiliation{California Institute for Quantitative Biosciences, University of California, Berkeley, CA 97420, USA}
\altaffiliation{Present address: Innovative Genomics Institute, University of California, Berkeley, CA 94720, USA}
\email{bibianaonoa@berkeley.edu}

\title[An \textsf{achemso} demo]
  {Focused helium ion and electron beam induced deposition of organometallic tips for dynamic AFM of biomolecules in liquid}

\begin{document}

%%%%%%%%%%%%%%%%%%%%%%%%%%%%%%%%%%%%%%%%%%%%%%%%%%%%%%%%%%%%%%%%%%%%%
%% The "tocentry" environment can be used to create an entry for the
%% graphical table of contents. It is given here as some journals
%% require that it is printed as part of the abstract page. It will
%% be automatically moved as appropriate.
%%%%%%%%%%%%%%%%%%%%%%%%%%%%%%%%%%%%%%%%%%%%%%%%%%%%%%%%%%%%%%%%%%%%%
%\begin{tocentry}

%Some journals require a graphical entry for the Table of Contents.
%This should be laid out ``print ready'' so that the sizing of the
%text is correct.

%Inside the \texttt{tocentry} environment, the font used is Helvetica
%8\,pt, as required by \emph{Journal of the American Chemical
%Society}.

%The surrounding frame is 9\,cm by 3.5\,cm, which is the maximum
%permitted for  \emph{Journal of the American Chemical Society}
%graphical table of content entries. The box will not resize if the
%content is too big: instead it will overflow the edge of the box.

%This box and the associated title will always be printed on a
%separate page at the end of the document.

%\end{tocentry}

%\linenumbers

%%%%%%%%%%%%%%%%%%%%%%%%%%%%%%%%%%%%%%%%%%%%%%%%%%%%%%%%%%%%%%%%%%%%%
%% The abstract environment will automatically gobble the contents
%% if an abstract is not used by the target journal.
%%%%%%%%%%%%%%%%%%%%%%%%%%%%%%%%%%%%%%%%%%%%%%%%%%%%%%%%%%%%%%%%%%%%%
\begin{strip}
\centering
\begin{minipage}{0.9\textwidth}

\begin{abstract}
  We demonstrate the fabrication of sharp nanopillars of high aspect ratio onto specialized atomic force microscopy (AFM) microcantilevers and their use for high-speed AFM of DNA and nucleoproteins in liquid. The fabrication technique uses localized charged-particle-induced deposition with either a focused beam of helium ions or electrons in a helium ion microscope (HIM) or scanning electron microscope (SEM). This approach enables customized growth onto delicate substrates with nanometer-scale placement precision and in-situ imaging of the final tip structures using the HIM or SEM. Tip radii of \unit[$<$10]{nm} are obtained and the underlying microcantilever remains intact. Instead of the more commonly used organic precursors employed for bio-AFM applications, we use an organometallic precursor (tungsten hexacarbonyl) resulting in tungsten-containing tips. Transmission electron microscopy reveals a thin layer of carbon on the tips. Consequently, the interaction of the new tips with biological specimens is likely very similar to that of standard carbonaceous tips, with the added benefit of robustness. A further advantage of the organometallic tips is that compared to carbonaceous tips they better withstand UV-ozone cleaning treatments to remove residual organic contaminants between experiments. Such contaminants are unavoidable during the scanning of soft biomolecules in liquid.
  Our tips can also be grown onto the blunted tips of previously-used cantilevers, thus providing a means to recycle specialized cantilevers and restore their performance to the original manufacturer specifications. Finally, a focused helium ion beam milling technique to reduce the tip radii and thus further improve lateral spatial resolution in the AFM scans is demonstrated.
  %250 words only - currently 250
\end{abstract}

\end{minipage}
\end{strip}

%%%%%%%%%%%%%%%%%%%%%%%%%%%%%%%%%%%%%%%%%%%%%%%%%%%%%%%%%%%%%%%%%%%%%
%% Start the main part of the manuscript here.
%%%%%%%%%%%%%%%%%%%%%%%%%%%%%%%%%%%%%%%%%%%%%%%%%%%%%%%%%%%%%%%%%%%%%
\newpage
\section{Introduction}

Dynamic atomic force microscopy (AFM) of biomolecules in liquid is concerned with directly observing the motion of molecules such as proteins and DNA with high spatiotemporal resolution under conditions closer to their native physiological environment~\cite{Ando2013,Ando2014}. These investigations are critical for understanding the basic mechanisms underlying a range of biological processes. In order to capture images of biomolecules in action, without interfering with their motion or suffering from motion-induced image blurring, minimally-invasive high-speed AFM (HS-AFM) techniques have been developed~\cite{Fantner2006,Brown2013,Ando2018,Fukuda2021}. In addition to various advances in scanning hardware and data acquisition systems, HS-AFM relies on miniaturized cantilevers that maximize the resonant frequency in liquid and minimize the spring constant in order to achieve the desired high-frame-rate and low-invasiveness performance. These cantilevers should be equipped with ultrasharp tips, since the tip radius determines the lateral spatial resolution of the AFM scanning technique. Furthermore, the tips should have a high aspect ratio, to enable nanoscale probing of steep topographic features~\cite{Ding2022}.

Various methods for creating sharp tips (or ``spikes") with high aspect ratio on AFM cantilevers have been explored, including focused ion beam milling~\cite{Ageev2015}, attachment of carbon nanotubes~\cite{Wilson2009}, in-situ growth of nanowires~\cite{Engstrom2011}, and focused electron/ion beam induced deposition~\cite{Plank2020,Ding2022}. For the growth of high aspect ratio tips on microcantilevers for HS-AFM of biomolecules in liquid, focused electron beam induced deposition (FEBID) using a carbonaceous precursor is typically used, which can be followed by a plasma-etching step in order to reduce the tip radius as needed~\cite{Uchihashi2012}. In the FEBID method, performed in a scanning electron microscope (SEM), the electron beam is focused onto a chosen location on the cantilever, which locally generates secondary electrons and backscatter electrons. These various electrons dissociate adsorbed precursor molecules and the non-volatile reaction products result in the growth of the tip~\cite{Utke2008}. In this way, the tip is grown directly onto the cantilever substrate with nanometer placement precision in a one-step process. The tip length is controlled by the electron dose of the primary beam. The precursor can be introduced in gaseous form via a gas injector needle, or using a vessel with a narrow outlet that is placed inside the vacuum chamber and contains a volatile compound. In fact in early work, residual hydrocarbons in the vacuum (\textit{e.g.}\ from pump oil) were used as the deposition source material~\cite{Akama1990}. In another method, the cantilever is first coated with a thin layer of hydrocarbons by immersion in oil solution, and the hydrocarbons are then locally polymerized by the electron beam to grow the tip~\cite{Walters1996}. 

An alternative localized deposition approach is focused ion beam induced deposition (FIBID)\cite{Utke2008,Orus2020}, typically using the focused ion beam (FIB) of a dual-beam FIB-SEM instrument. Similar to FEBID, various beam-sample interactions are responsible for the dissociation of the precursor molecules. However, generally the deposition rates of FIBID are significantly higher than those of FEBID~\cite{De_Teresa2009}, attributed to the higher secondary electron emission yields and the additional contribution of excited surface atoms~\cite{Utke2008}. To date, most FIBID has been performed using the conventional gallium FIB. However, for thin and delicate substrates such as the HS-AFM cantilevers of interest in the present work, the gallium FIB method is problematic, since the gallium ions erode the substrate by sputtering before the deposit can form. This is where the more-recently introduced helium FIB brings benefit\cite{Allen2021b}. The light helium ions have a much lower sputter yield and as we show here, leave the cantilever intact. Helium-FIBID tips are much narrower than their gallium-FIBID counterparts, due to the smaller probe size and narrower interaction volume of the helium ions near the surface\cite{Alkemade2014}. Over the last few years, the use of helium-FIBID to fabricate high aspect ratio nanostructures has been increasingly demonstrated, creating \textit{e.g.}\ hollow\cite{Cordoba2018} and helical\cite{Cordoba2019} nanowires, and hammerhead-shaped AFM probes\cite{Nanda2015}.

In this work we use both FEBID and helium-FIBID to fabricate nanopillars onto specialized HS-AFM microcantilevers and employ them to scan DNA and nucleoproteins in liquid at high frame rates. The lengths of our tips are \unit[200--600]{nm} with tip radii of \unit[$<$10]{nm}. In addition, we show that the FEBID and helium-FIBID methods can be used to grow new sharp tips onto the blunted tips of previously-used cantilevers, thus enabling re-deployment of specialized HS-AFM cantilevers over many cycles. The resolution performance of our tips equals that of the commercial tips we compare them with and we demonstrate how helium-FIB milling can be used to create even sharper tips to further increase the lateral spatial resolution of the AFM scans.

Rather than using one of the carbonaceous precursors more typically used for bio-AFM applications, we use an organometallic precursor containing tungsten. Compositional analysis reveals that the outer surface of the resulting organometallic tips is composed of a thin carbonaceous layer, hence the tip--sample interaction is expected to be comparable to that of tips grown from carbon-based precursors. The particular benefit of the organometallic tip in our case is that it renders the tips much more robust to cleaning treatments between experiments. During these treatments, only surface contaminants should be removed and the tip itself should remain intact; this is much more difficult to achieve if the tip is also carbon-based. A wide range of metal-containing precursor chemistries is available. Moreover, as has been shown in a recent FEBID study, iron-containing precursors can be used to grow magnetic tips onto HS-AFM cantilevers and employed for magnetic force microscopy (MFM) in liquid~\cite{Jaafar2020}. Such tips could also be used for in-liquid MFM of magnetic biological structures~\cite{Winkler2023}, further underscoring the versatility of the FEBID and FIBID tip fabrication methods.

\section{Materials and Methods}

\subsection{Organometallic tips grown by He-FIBID}

Deposition by He-FIBID was performed using a Zeiss ORION NanoFab helium ion microscope (HIM). The gas field ionization source (GFIS) was operated with a helium gas pressure of \unit[2~$\times$~10$^{-6}$]{Torr} at an acceleration potential of \unit[25]{kV}. Selecting a \unit[10]{{\textmu}m} beam-limiting aperture gave a probe current of \unit[1--2]{pA} (nominal probe size of \unit[$\sim$0.5]{nm}, spot control value set to 4). The working distance was set to \unit[9]{mm}. 

The organometallic gaseous precursor tungsten hexacarbonyl (W(CO)$_6$) was introduced via a needle placed \unit[$\sim$100]{{\textmu}m} from the target region on the sample using an Oxford Instruments OmniGIS II gas injector system (cartridge temperature \unit[50]{$^{\circ}$C}). The chamber pressure during flow of the precursor gas was \unit[8$\times$10$^{-6}$]{Torr} (chamber base pressure \unit[2$\times$10$^{-7}$]{Torr}). NanoPatterning and Visualization Engine (NPVE) software from Fibics, Inc. was used to control the beam during the FIBID process, selecting continuous dwell `spot mode' exposures to grow the nanopillars. Before each nanopillar deposition, care was taken to achieve the best beam focus (smallest spot size) and to correct for any beam stigmation. Typically this involved depositing shorter test nanopillars onto the wider base regions of the HS-AFM cantilevers. The He-FIBID deposition times ranged from \unit[2--6]{s} producing tips of length \unit[200--600]{nm}.

Two types of HS-AFM microcantilevers were used: Nanoworld USC-F1.2-K0.15-10 with cone-shaped tips and Olympus BL-AC10DS-A2 with ``bird-beak'' style tips, from here on named Type I and Type II, respectively. In all cases, the cantilevers were mounted horizontally using double-sided conductive carbon tape (3M\textsuperscript{TM} XYZ Axis Tape) onto a SEM stub. For deposition onto Type I tips, the microscope stage was kept at or close to \unit[0]{$^{\circ}$} tilt so that the central axis of the nanopillar deposit was parallel to the central axis of the cantilever cone (the cones were first imaged end-on with the HIM (vertical column) and slight tilt/rotation adjustments were made as needed). For the Type II cantilevers, the stage was tilted by \unit[10]{$^{\circ}$} so that in this case, the angle between the cantilever plane and the new tip was \unit[100]{$^{\circ}$}\cite{Uchihashi2012}. After the depositions, low-dose images of the deposits were acquired using HIM. 

A separate set of He-FIBID deposits for high-resolution inspection and composition analysis (see below) was also grown onto a copper TEM half-grid following a protocol introduced previously\cite{Allen2021}.

\subsection{Organometallic tips grown by FEBID}

The FEBID deposition was performed using an FEI (now Thermo Fisher) Helios NanoLab 600 FIB-SEM instrument. The acceleration potential of the SEM column was set to \unit[30]{kV} and a beam current of \unit[21]{pA} was selected. 

The same organometallic gaseous precursor chemistry (W(CO)$_6$) was used as for the He-FIBID, again delivered via a needle placed \unit[$\sim$100]{{\textmu}m} from the target region (cartridge temperature ranged from \unit[50--65]{$^{\circ}$C}). The chamber pressure during flow of the precursor gas was \unit[2.7$\times$10$^{-5}$]{Torr} (chamber base pressure \unit[1.6$\times$10$^{-6}$]{Torr}).
The FEBID tips were grown by implementing a circular scan of small radius (\unit[1 or 10]{nm}), controlling the final height of the tip with the number of circular passes (10,000--40,000). Deposition times varied from \unit[20--40]{s}, achieving tips that were \unit[300-600]{nm} in length. The same cantilever mounting and stage tilting protocols were implemented as for the He-FIBID tips. SEM images of the deposits were acquired at \unit[5]{kV}.

Finally, a set of FEBID deposits was also grown on a copper TEM half-grid for high-resolution analysis.

\subsection{(S)TEM analysis of tips}

Bright-field TEM imaging, dark-field scanning TEM (STEM) imaging, and elemental mapping by STEM-based X-ray energy-dispersive spectrometry (XEDS) were performed using an FEI TitanX electron microscope operated at \unit[300]{kV}. The XEDS maps were acquired using an FEI Super-X quadrature X-ray detector and Bruker Esprit software.

\subsection{High-speed AFM of bio-molecules in liquid}

The fabricated He-FIBID and FEBID tips were used to image both individual DNA molecules and nucleoproteins in buffer. The DNA samples were either linearized pUC19 DNA plasmid (\unit[1,982]{bp}) or a DNA fragment (\unit[660]{bp}) diluted at \unit[2]{nM} and \unit[0.5]{nM}, respectively, in deposition buffer (\unit[10]{mM} MOPS pH 7.0 and \unit[2]{mM} MgCl$_2$)). The nucleoprotein samples (nucleosomes or tetrasomes) were diluted at \unit[3]{nM} in buffer A (\unit[5]{mM} MOPS pH 7.0, \unit[80]{mM} KCl, \unit[20]{mM} NaCl, \unit[0.5]{mM} EGTA, \unit[2]{mM} EDTA, \unit[0.5]{mM} spermidine, \unit[0.2]{mM} spermine, \unit[5]{mM} Na(C$_3$H$_7$COO), \unit[1]{mM} DTT)). The diluted samples were deposited onto freshly cleaved mica and left for \unit[2]{min}. The surface of the substrate was then rinsed to remove unbound species and the samples were imaged in buffer (\unit[10]{mM} MOPS pH 7.0 for the DNA, and 10-fold diluted buffer A for nucleoproteins).

The AFM used was an Ando model HS-AFM (Research Institute of Biomolecule Metrology Co., Ltd.\ Japan). In order to avoid overlapping molecules, one molecule was scanned at a time. All samples were scanned in liquid (buffer composition is specific for each sample, as outlined above) using tapping mode at room temperature. A range of scan rates was implemented. The spring constants of the Type I (cone style) and Type II (bird-beak style) cantilevers were \unit[$\sim$0.1]{N/m} and the resonance frequency in liquid \unit[600]{kHz}. 
The deflection of the cantilever was detected using an optical beam detector. 
A0 was set to \unit[$\sim$2]{nm}, and the As/A0 ratio was kept high to achieve the highest resolution with the lowest force possible. When necessary, used probes were cleaned by UV-ozone oxidation by irradiating the cantilever locally using a pencil style Hg(Ar) UV lamp (Oriel Mod. 6035 pencil-style spectral calibration lamp from Newport Inc.) for \unit[1--3]{min}.

\section{Results and discussion}

\subsection{Structure and composition analysis of the organometallic tips}

\begin{figure}[t]
\includegraphics[width=0.95\linewidth]{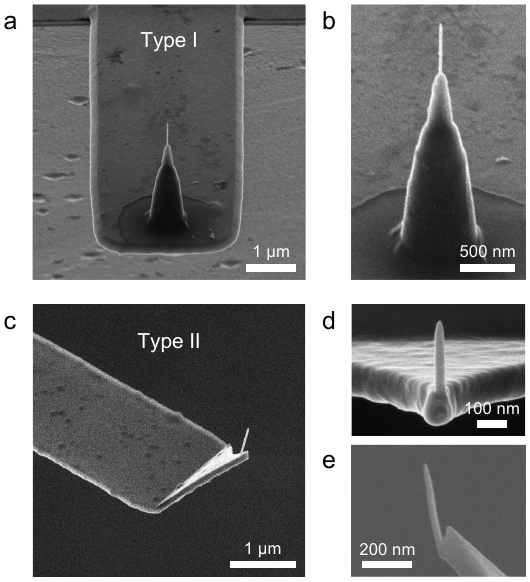}
  \caption{HIM/SEM images showing the two HS-AFM cantilever types and examples of the He-FIBID/FEBID tips grown for this study: (a) Cone-shaped cantilever (Type I), and (b) High-magnification view of He-FIBID tip grown at the apex; (c) Bird-beak-style cantilever (Type II), and (d) High-magnification view of He-FIBID tip grown at the apex, and (e) FEBID tip grown at the apex.}
  \label{fgr:1}
\end{figure}

Representative images of the He-FIBID and FEBID tips that were grown onto two types of HS-AFM microcantilever are shown in Figure~\ref{fgr:1}. The He-FIBID tips were imaged by helium ion microscopy (HIM) and the FEBID tips were imaged by SEM. The two cantilever types had either cone-shaped tips (named here Type I) or bird-beak-style tips (Type II), onto which our FIBID/FEBID tips were grown. A Type I cantilever with an He-FIBID tip is shown in Fig.~\ref{fgr:1}(a), together with a higher-magnification view in Fig.~\ref{fgr:1}(b). (The cone-shaped structure on the Type I cantilever onto which the new tip was grown had been blunted from prior use.) A Type II cantilever, also with an He-FIBID tip, is shown in Fig.~\ref{fgr:1}(c), with a higher-magnification view in Fig.~\ref{fgr:1}(d). A FEBID tip deposited onto a Type II cantilever is shown in Fig.~\ref{fgr:1}(e).

The images in Fig.~\ref{fgr:1} demonstrate the high placement precision of the He-FIBID and FEBID growth processes and the high aspect ratio of the resulting tips. Tip lengths were up to \unit[600]{nm} and tip base diameters were around \unit[30--50]{nm}.

\begin{figure}[ht]
\includegraphics[width=0.9\linewidth]{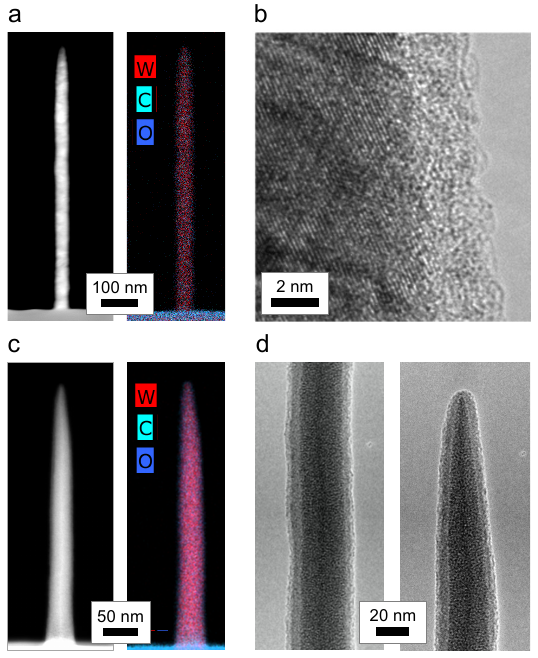}
  \caption{High-resolution (S)TEM analysis of He-FIBID and FEBID tips: (a) Dark-field STEM (left) and STEM-XEDS composite elemental map (right) for He-FIBID tip; (b) Bright-field TEM of surface region. (c) Dark-field STEM (left) and composite STEM-XEDS elemental map (right) for FEBID tip; (d) Bright-field TEM of surface regions.}
  \label{fgr:2}
\end{figure}

In order to investigate the structure and composition of the tips in greater detail, analysis by (scanning) transmission electron microscopy ((S)TEM) was performed, as summarized in Fig.~\ref{fgr:2}. In Fig.~\ref{fgr:2}(a), elemental mapping of an He-FIBID tip by STEM-based X-ray energy dispersive spectroscopy (XEDS) reveals that the tip is composed of W, C and O. This is expected, since the organometallic gaseous precursor used for the deposition was W(CO)$_6$, which contains all these elements. The implantation of helium ions into the material during the growth process is also expected, but this could not be mapped as there is no characteristic X-ray for He. We speculate that in the case of the narrow nanopillar structure, implanted helium atoms could have exited by diffusion to the surface. In any event, certainly no high-dose helium irradiation effects were observed, such as the formation of helium nanobubbles, as has been reported for high-dose helium irradiation of bulk tungsten targets \cite{Allen2020}, or the formation of a hollow core inside the tip due to concurrent milling by the focused helium ion beam\cite{Cordoba2018}. The fact that no hollow cores were formed in our He-FIBID tips points to fast deposition rates that strongly outcompeted the milling rate.

Figure~\ref{fgr:2}(b) shows a high-magnification TEM view of the sidewall of the He-FIBID tip and reveals a key feature, namely an amorphous surface layer of thickness \unit[$\sim$3]{nm}. Inspection of the individual elemental maps for W, C, and O reveals that the surface layer is composed of C and O (see Supplementary Fig.~S1). 
This is significant, because it means that the metal component of the tip (W) will not come into direct contact with the biomolecules in the HS-AFM experiments, as will be discussed further below. The carbonaceous surface layer may already form inside the HIM instrument directly after/during the deposition due to residual hydrocarbons on the sample and/or in the sample chamber\cite{Hlawacek2013}. It is also well-known that samples contaminate quickly with a thin layer of hydrocarbons upon exposure to air~\cite{Barr1995}.

Turning to the FEBID tips, Fig.~\ref{fgr:2}(c) shows STEM-XEDS elemental mapping results for one such tip. Since the same W(CO)$_6$ precursor chemistry was used, the FEBID tip was again found to comprise W, C and O. The relative amounts of these elements in the He-FIBID versus FEBID tips did, however, vary. For the tips analyzed here, the relative composition of the He-FIBID tip material was: \unit[58]{at.\%} W, \unit[26]{at.\%} C, \unit[16]{at.\%} O, whereas that of the FEBID tip material was \unit[22]{at.\%} W, \unit[45]{at.\%} C, \unit[33]{at.\%} O. It is generally understood that the metal fraction of FIBID tips is higher than that of FEBID tips deposited from the same organometallic precursor, due to more efficient dissociation of the precursor by the ions~\cite{De_Teresa2009}. This means that the amorphous component of the central FEBID material will be higher and its crystalline component will in turn be lower, with potential implications for the mechanical properties of the tips\cite{Utke2020}. However, in the present imaging study, no discernible difference between the performance and longevity of the He-FIBID \textit{vs.}\ FEBID tips could be determined.

As was the case for the He-FIBID tips, TEM analysis of the FEBID tips also revealed an amorphous suface layer, as seen in Fig.~\ref{fgr:2}(d). The thickness of the surface layer varied between individual FEBID tips and was generally thicker than for the He-FIBID tips (up to \unit[$\sim$10]{nm} in thickness). These variations may be influenced by fluctuations in precursor flow rate, differences in the chamber base vacuum between the two instruments used for the He-FIBID \textit{vs.}\ FEBID, and the fact that the FEBID tips traveled (in air) from Spain to California before the TEM analysis. Regardless, for the HS-AFM the end result is the same, \textit{i.e.}\ the metallic component of the FEBID tip will not come into direct contact with the HS-AFM samples.

\subsection{HS-AFM of DNA using the organometallic tips}

\subsubsection{First tests}

Typical cantilever tips used for AFM of biological specimens are carbon- or silicon-based, hence the first question to address was whether the metal-containing tips fabricated in this study are also suitable for AFM imaging of biological samples. Specifically, the probe--sample interactions should achieve attractive/repulsive forces within a range suitable to obtain stable and high signal-to-noise ratio images without damaging the soft biomolecules. The metal in question here is tungsten (from the W(CO)$_6$ precursor). Tungsten is a heavy rare metal and its compounds are generally inert~\cite{Strigul2005}. However, tungsten is also known to be biologically functional, being present in oxidoreductases and in non-redox enzymes that play an important metabolic role in all living things (\textit{e.g.}\ as nitrate reductase, sulfite oxidase, acetylene hydratase, \textit{etc.}).~\cite{Seiffert2007}. It has also been reported that tungsten carbide in a cobalt matrix can damage DNA via reductive oxygen species~\cite{Kanellis2018}. To test our tungsten-containing AFM tips, we thus used a range of scanning and biologically-relevant conditions and compared the results with those obtained using commercial high-density carbon tips.

\begin{figure}[t]
\includegraphics[width=0.9\linewidth]{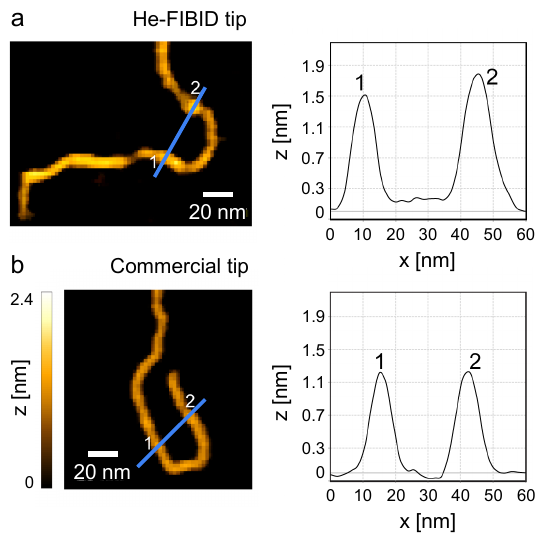}
  \caption{HS-AFM results for DNA (\unit[660]{bp}) fragments scanned at \unit[1]{f/s} using (a) He-FIBID tip on recycled Type I cantilever (see Fig.\ \ref{fgr:1}(a)), compared to (b) Type I cantilever with original high-density carbon cone-shaped tip. The color bar corresponds to the heights measured in both images. The adjacent plots show line profiles across the DNA loops (corresponding to the blue lines in the images) enabling analysis of the DNA height and width profiles for the points labeled 1 and 2.}
  \label{fgr:3}
\end{figure} 

In Fig.~\ref{fgr:3}, we compare HS-AFM data obtained using a tungsten-based He-FIBID tip and a commercial high-density carbon probe. The test sample comprised DNA fragments in buffer scanned at \unit[1]{f/s} and the cantilever type employed was the Type I variety (see Fig.\ \ref{fgr:1}(a)). In the He-FIBID case, shown in Fig.~\ref{fgr:3}(a), a used cantilever that had a blunted carbon cone-shaped tip served as the base onto which the organometallic tip was grown (thus effectively recycling the old probe). The control probe, giving the results in Fig.~\ref{fgr:3}(b), was a new and unmodified cone-shaped carbon probe. To summarize the result of the comparison, we find that the organometallic He-FIBID and high-density carbon commericial probe produced comparable HS-AFM results, indicating that the composition of the organometallic tips fabricated for this study does not appear to modify the probe--sample interaction (the force-distance curve measurements performed on clean mica prior to DNA deposition were also comparable). As will be seen, the organometallic FEBID tips also produced similar results. On the basis of the tip composition analysis shown in Fig.~\ref{fgr:2}, we conclude that the carbonaceous shell on the He-FIBID and FEBID tips is responsible for the apparent similarity in probe--sample interaction compared to the carbon-based commercial probes. The apex diameters of our tips are also comparable, as we now discuss. 

DNA is a double helix of diameter \unit[$\sim$2]{nm}~\cite{Drew1981}, thus the height of the DNA molecule fragments as imaged by AFM in liquid is expected to be close to this value when using tapping mode at low amplitudes (\textit{i.e.}\ low force). In the line profile measurements of Fig.~\ref{fgr:3} (corresponding to the blue lines in the AFM images), we see that we are close to this expected height value, with as-scanned heights of the labeled points on the DNA loop ranging from \unit[1.2 to 1.8]{nm}. The as-scanned widths of the DNA molecule are larger, since this dimension is ultimately determined by the radius of curvature of the probe apex~\cite{Martinez2011}. For the He-FIBID and the commercial probe, full width at half maximum (FWHM) values corresponding to these width measurements range from \unit[8.2 to 9.8]{nm}, indicating a radius of curvature of the probes of \unit[$\sim$3--4]{nm}. In all cases, the height and width measurements obtained using the He-FIBID and commercial tips are comparable and in agreement with literature reported values~\cite{Lyubchenko2012}. Due to the good agreement, tip deconvolution analysis from the AFM line profiles was not performed. The complete set of measured FWHM and height values from the HS-AFM data sets is given in Supplementary Table~S1. 

\begin{figure*}[t]
\includegraphics[width=0.8\linewidth]{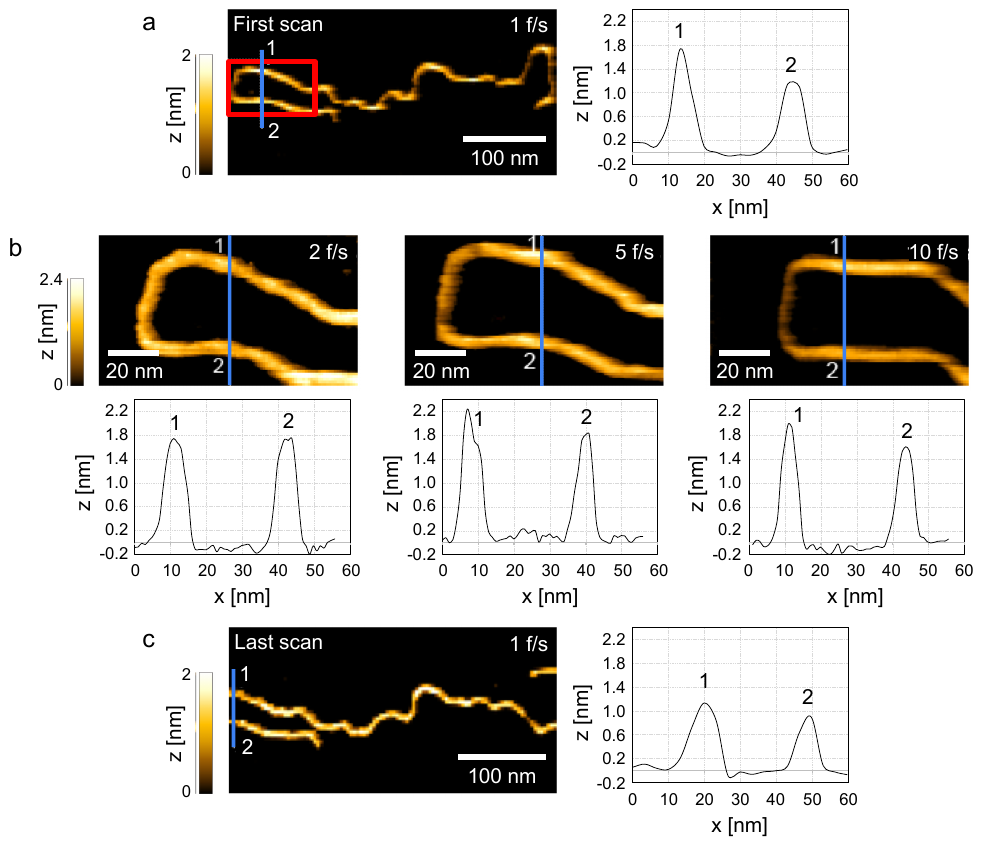}
  \caption{HS-AFM results for a long DNA molecule (\unit[1,982]{bp}) scanned with a FEBID tip on a Type II cantilever (see Fig.~\ref{fgr:1}(e)) probing continuous scanning for a range of scan rates and fields of view. (a) Large field-of-view scan at \unit[1]{f/s} showing the full length of the DNA molecule at the beginning of the test. (b) High-resolution scans of DNA loop portion (red box in (a)) for \unit[2, 5, 10]{f/s} scan rates. (c) Large field-of-view scan at \unit[1]{f/s} showing (almost) the full length of the DNA molecule at the end of the test. Each AFM image corresponds to the first scan from a \unit[45--60]{s} movie. Line profiles (blue lines) across the DNA loop track the as-measured height and FWHM values of the points labeled 1 and 2.}
  \label{fgr:4}
\end{figure*}

\subsubsection{Testing different scan rates}

In order to test the overall performance of the tips, a long DNA molecule (\unit[1,982]{bp}) was scanned repeatedly using various scan rates, as summarized in Fig.~\ref{fgr:4}. The tip used in this example was a FEBID tip on a Type II cantilever (see Fig.~\ref{fgr:1}(e)). This test required that the DNA was strongly bound to the substrate surface, which was achieved using Mg$^{2+}$ ions during the sample deposition. Figure~\ref{fgr:4}(a) shows the first scan, corresponding to the first frame of a movie of the full length of the DNA molecule acquired at \unit[1]{f/s} for \unit[$\sim$60]{s}. The region highlighted with the red box (looped part of the molecule) was then scanned at higher resolution at \unit[2, 5, and 10]{f/s} for \unit[45--60]{s} at each frame rate; the first scan from each of these movies is shown in Fig.~\ref{fgr:4}(b). This set of high-resolution scans at the three different scan rates was then repeated and finally, the full molecule was scanned again at \unit[1]{f/s} for \unit[$\sim$60]{s}. The first scan from the final movie is shown in Fig.~\ref{fgr:4}(c).

\begin{figure*}[]
\includegraphics[width=0.95\linewidth]{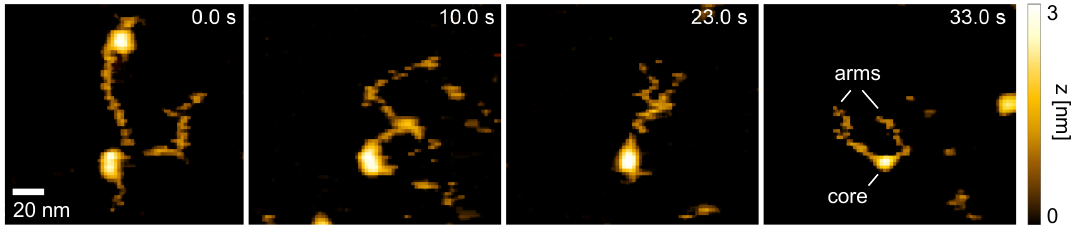}
  \caption{HS-AFM time lapse of mobile nucleoprotein tetrasomes recorded at \unit[1]{f/s} using He-FIBID tip on recycled Type II cantilever.}
  \label{fgr:5}
\end{figure*}

The line profile plots in Fig.~\ref{fgr:4} correspond to the blue lines marked in the images and track two points (labeled 1 and 2) on the looped part of the molecule. In any given scan, the as-measured height of the DNA molecule fluctuates (as also seen in Fig.~\ref{fgr:3}). Such intramolecular fluctuation is expected, since inhomogeneous adsorption of the molecule on the charged surface will induce local mechanical stress, resulting in variations in the final 2D conformation of the double helix along its length~\cite{Prokhorov2021}. Regardless, the molecule heights measured using the FEBID tip shown in Fig.~\ref{fgr:4} are comparable to those measured using the He-FIBID and unmodified commercial carbon-based probe shown in Fig.~\ref{fgr:3}. 
The full set of measured values (heights and FWHM) from the line profiles of Fig.~\ref{fgr:4} are given in Supplementary Table~S2.

Due to the mobility of the DNA molecule (lateral as well as vertical), it is difficult to draw conclusions from the FWHM values. However, the line profile results suggest that scanning at the faster frame rates (\unit[$\ge$5]{f/s}) yields DNA dimensions closer to the expected values, both in terms of height (close to the expected DNA width, \textit{i.e.}~\unit[$\sim$2]{nm}) and in terms of yielding the smallest FWHM values (down to \unit[4.4]{nm}). Thus the higher scan rates appear to capture the movement of the DNA molecule more effectively. Overall, the results demonstrate that the high spatiotemporal resolution of the FEBID tip was preserved after continuous scanning for a range of scan rates.

\subsection{HS-AFM of nucleoproteins using the organometallic tips}

We further challenged our He-FIBID/FEBID probes by scanning loosely bound nucleoproteins. For example, Fig.~\ref{fgr:5} shows HS-AFM time lapse results for tetrasome molecules in motion, acquired with an He-FIBID tip that had been grown on a Type II cantilever tip (recycled). 

A tetrasome is a complex of two histone heterodimers arranged such that there is a protein core (comprising four histones) with two DNA `arms'. While various experimental protocols have been developed to purposefully mediate the adhesion of DNA onto the mica surface (both the DNA and the surface are negatively charged), our experiment was set up specifically to allow motion of the DNA; \textit{i.e.}\ no functionalization of the mica surface was performed and no divalent cations were included in the buffer to mediate binding~\cite{Japaridze2016,Heenan2019}. Consequently, our image series captures the DNA arms of the tetrasomes in motion as they transiently desorb and readsorb onto the mica. When the DNA filaments desorb into solution, they temporarily become invisible, which is reflected by the discontinuous nature of the filaments in the images. In contrast to the DNA, the histone core of each tetrasome has a net positive charge and thus experiences an electrostatic attraction to the negatively charged mica. Nevertheless, our results show that the histone cores also moved around and we observe tetrasomes interacting with one another and diffusing out of the field of view. All this was captured at high spatiotemporal resolution using the He-FIBID tip.

\subsection{Tip endurance test}

\begin{figure}[]
\includegraphics[width=0.9\linewidth]{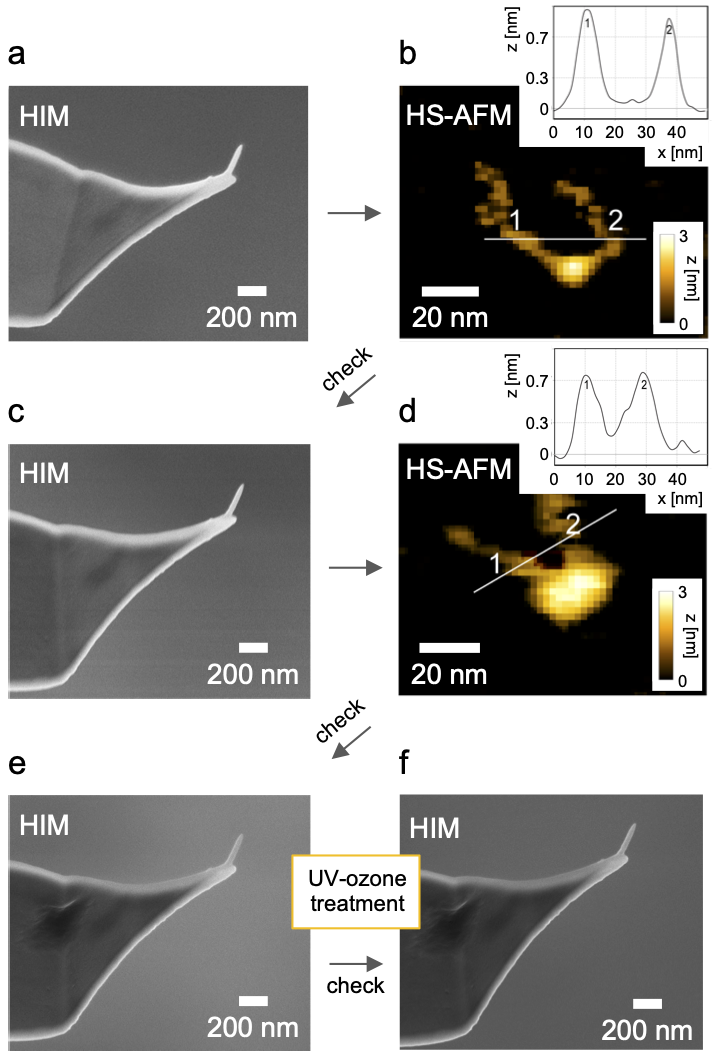}
  \caption{Results from a tip endurance test in which an He-FIBID tip on a Type II cantilever was used to scan nucleoprotein tetrasomes and subjected to a UV-ozone cleaning treatment. (a) HIM image of tip before scans. (b) HS-AFM image of tetrasome with line profile analysis from first round of AFM scans. (c) Subsequent HIM image of tip. (d) HS-AFM image from second round of AFM scans with line profile analysis. (e) Subsequent HIM image of tip. (f) Final HIM image of tip following UV-ozone treatment.}
  \label{fgr:6}
\end{figure}

Usually, if a tip underperforms in HS-AFM experiments (\textit{i.e.}\ yields poor spatial resolution, artifacts, \textit{etc.}), it is either discarded, or if contamination is suspected, attempts are made to clean the tip. In our work, we specifically wanted to investigate the robustness of our tips and also check the effect of UV-ozone cleaning on the tip condition. Therefore, a series of tests was performed where a tip was imaged by HIM directly before and after 1) separate HS-AFM scanning experiments, and 2) a UV-ozone cleaning treatment, as summarized in Fig.~\ref{fgr:6}. 

First, an He-FIBID tip on a Type II cantilever was imaged by HIM immediately after its fabrication (Fig.~\ref{fgr:6}(a)). Next the tip was used to image a loosely bound tetrasome sample (Fig.~\ref{fgr:6}(b)). The HS-AFM image shown here is from the time series shown in Fig.~\ref{fgr:5}; line profile analysis is also shown. After continuous scanning at \unit[1]{f/s}, the tip was removed from the solution, dried in air, and re-imaged by HIM (Fig.~\ref{fgr:6}(c)). The tip was then used to scan another (fresh) tetrasome sample (Fig.~\ref{fgr:6}(d)), after which it was again inspected by HIM (Fig.~\ref{fgr:6}(e)). In the final step, the tip was subjected to a dry UV-ozone treatment (see Methods) to test the effect of a simple method to remove organic contaminants from non-carbon-based probes (\textit{e.g.}\ metal-coated or silicon tips)~\cite{Kohli2019}. The HIM image of the tip after UV-ozone treatment is shown in Fig.~\ref{fgr:6}(f).

The first key finding from the set of experiments outlined in Fig.~\ref{fgr:6} is that the He-FIBID tip survived the rounds of HS-AFM as well as the UV-ozone treatment. The weak point in the architecture of the tip is expected to be near or at its base, where it connects to the cantilever. Indeed, in a previous experiment it was found that an He-FIBID tip had fractured near the base, which either occurred during the HS-AFM scanning, or upon transfer of the cantilever out of the HS-AFM setup. Novel architectures for FEBID tips have been explored, incorporating broader bases or tripod-like structures to increase stability~\cite{Plank2020,Seewald2022}.
While such approaches were not investigated in the present work, they could be explored in the future. Nevertheless, Fig.~\ref{fgr:6} shows that if the tips are handled carefully, they are robust enough for repeated use. 

To probe the endurance of the tip apex, we compare the line profile analysis of the nucleoprotein DNA `arms' from the HS-AFM scans of Figs.~\ref{fgr:6}(b) and (d). We observe that the scanned heights decreased slightly while the FWHM values increased slightly (see Supplementary Table~S3). This could indicate slight broadening of the tip apex, which is possibly observed in the HIM image of Fig.~\ref{fgr:6}(e). It should be noted, however, that the mobility of the specimen may also be contributing to the apparent broadening in the HS-AFM scans. In any case, the HS-AFM results from Fig.~\ref{fgr:6}(d) show that the tip is still able to clearly resolve the dynamics of the tetrasome samples, showing the DNA `arms' as discontinuous filaments due to their dynamics and well-resolved histone cores, which are less mobile, as described previously. 

With samples such as these, contamination of the probe is to be expected. For example, during DNA desorption, filaments interacting with organic debris can make contact with the tip and transfer to the tip apex. Indeed, it is well known that contamination with organic materials during HS-AFM of biomolecules in liquid can cause loss of spatial resolution and general image deterioration due to tip broadening and/or ghost imaging (the latter resulting from so-called double tips). 
While such severe effects were not experienced here, and no obvious contamination on the tips was observed (at least not at the magnification of the HIM images), the possibility of contamination buildup still needs to be considered. This was the motivation for investigating the UV-ozone cleaning step. Ozone treatment relies on the selective oxidation of biomolecules, which is exploited broadly in industry for decontamination, storage and disinfection purposes~\cite{Epelle2023}. However, carbon-based AFM tips can potentially also be etched and damaged upon ozone exposure. In contrast, the tungsten-based tips fabricated here consistently survive the treatment, as can be seen when comparing Figs.~\ref{fgr:6}(e) and (f). Elsewhere in this study, the UV-ozone treatment was also routinely employed to clean the FEBID/He-FIBID tungsten-based tips between scans, as needed (\textit{e.g.}\ for the experiments summarized in Fig.~\ref{fgr:4}, testing different scan rates over many cycles with a single FEBID tip). No change in tip--sample interaction after cleaning was observed, hence we conclude that the surface hydrocarbon layer either survived the ozone treatment, or a new hydrocarbon layer soon formed upon exposure to air due to so-called adventitious carbon~\cite{Barr1995}. Our results thus indicate that by choosing a metal-based tip, a simple and affordable UV-based cleaning strategy can be employed to remove organic contaminants and thus extend tip lifetimes for future experiments.

\subsection{Strategy to further decrease tip radius}

A persistent need in all areas of microscopy is to further increase spatial resolution. For AFM, this can be achieved by decreasing the radius of curvature of the apex of the cantilever tip. The tips fabricated by He-FIBID and FEBID in this work already have tip radii in the range of \unit[4--10]{nm} (inferred from the high-resolution TEM images, HIM images, and the HS-AFM line profile analysis for the DNA molecule). These tip radii are comparable to those of high-end commercial HS-AFM probes. 

\begin{figure}[t]
\includegraphics[width=\linewidth]{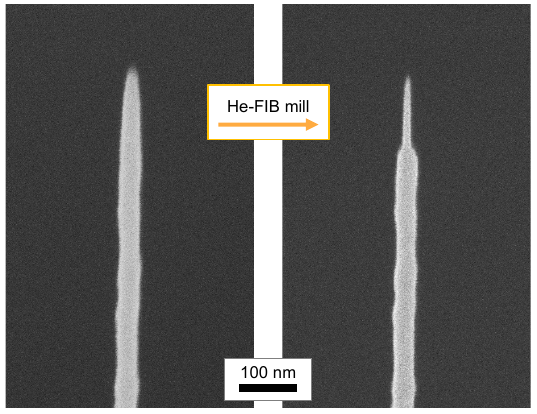}
  \caption{HIM images showing He-FIBID tip that has been sharpened by He-FIB milling.}
  \label{fgr:7}
\end{figure}

In order to decrease the tip radii even further, the He-FIB method offers an enticing opportunity, since as well as inducing deposition, the focused helium ion beam can also be used for precise milling of features down to the single-digit nanometer scale~\cite{Allen2021b}. To test this, He-FIB milling was used to sharpen an He-FIBID tip, as shown in Fig.~\ref{fgr:7}. The milling strategy employed here was to mill from the side using rectangular patterns placed over opposite edges of the tip body in the apex region (see Supplementary Fig.~S2). This resulted in unidirectional thinning, bringing the tip radius in the HIM imaging plane down from \unit[$\sim$8 to 3]{nm}. The tip can be rotated to allow thinning from other directions as well. Top-down annular milling strategies, as are applied in FIB-milling of needle-shaped specimens for atom probe tomography could also be explored~\cite{Allen2023}, but milling of the underlying cantilever would need to be kept to a negligible amount. 

Compared with other tip sharpening techniques, such as plasma treatment~\cite{Wendel1995}, thermal annealing combined with reactive ion etching~\cite{Kometani2021}, or electron beam induced purification in the presence of atmospheric water molecules~\cite{Plank2020}, the advantage of localized He-FIB milling from the side (as demonstrated here) is that the delicate HS-AFM microcantilever is not impacted. However, we note that the resulting ultrasharp tips need to be used with extra care, since soft biological structures can be punctured and more easily damaged during scans.

\section{Conclusions}

In this study we have shown that focused beams of helium ions or electrons can be used for localized beam induced deposition to fabricate robust high aspect ratio tips with tip radii \unit[$<$10]{nm} for high-speed AFM of biomolecules in liquid. While the electron beam method (FEBID) has been demonstrated widely for AFM tip fabrication, the much more recent He-FIBID method is still underexplored in this area. Compared with more conventional FIBID using gallium ions, He-FIBID has the distinct benefit of 1) enabling deposition onto delicate substrates, such as the \unit[$\sim$150]{nm} thick bird-beak style microcantilevers, without erosion of that substrate, and 2) producing narrower tips due to the narrower interaction volume of the light ions close to the surface. We also show how targeted milling with the He-FIB can be used to thin tips from the side to reduce tip radii to just a few nanometers. 

Another key aspect of this study is that the tips were fabricated using an organometallic precursor (tungsten hexacarbonyl) rather than one of the conventional carbon-based precursors that are usually used to fabricate tips for biological AFM. We show that even though our tips contain tungsten, which is known to be biologically functional, the tip--sample interactions appear to be essentially the same as for carbon-based probes. This is attributed to a thin hydrocarbon layer (several nanometers) coating the tungsten-rich core of the tip, which we imaged by TEM. 

The He-FIBID and FEBID tips were tested by imaging DNA and nucleoproteins in liquid, implementing scan rates from \unit[1--10]{f/s}. The performance of both tip types compared well with high-end electron-beam-deposited carbon-based commercial probes. Conveniently, the He-FIBID/FEBID tip fabrication methods work equally well on new tip-less cantilevers as well as on used cantilevers with damaged/worn tips. Comparing the lengths of the various He-FIBID/FEBID tips tested (\unit[$\sim$200--600]{nm}), no effect on AFM image performance or quality was observed. Tip length is easily tunable by the deposition dose (or equivalently the deposition time)~\cite{Allen2021}. In fact, He-FIBID tips with lengths of several microns have been reported in several works~\cite{Hill2011,Cordoba2018,Allen2021}. 

The customized He-FIBID/FEBID tips successfully enabled continuous imaging of DNA and nucleoproteins in buffer at different scan rates for several hours without introducing undesired molecular distortions or scanning artifacts. Furthermore, the lifetime of the organometallic tips after prolonged use can be extended using a simple UV-ozone cleaning treatment. In the case of carbon-based tips, UV-ozone treatment has the potential to remove both the organic contaminants as well as etch the tip itself, hence must be delicately controlled. In contrast, the organometallic tips fabricated here are by nature much more robust to the cleaning method and thus a single tip can be employed for various HS-AFM experiments scanning different samples, with UV-ozone treatments in between. Even if the UV-ozone treatment removes the thin carbon layer from the organometallic tips (to which we attribute the favorable tip--sample interactions), adventitious carbon is likely to quickly redeposit upon exposure to air~\cite{Barr1995}.
Once a tip is finally exhausted, a new one can be fabricated at the desired location on the same cantilever using the He-FIBID/FEBID procedure. In future work, further customization of the tips can be explored, including the fabrication of wider tip bases~\cite{Plank2020,Seewald2022} and fine-tuning tip compositions~\cite{Utke2020} for enhanced mechanical stability in broader AFM user environments including for conductivity and magnetic measurements.

\clearpage
\begin{acknowledgement}

The authors thank the Biomolecular Nanotechnology Center of QB3-Berkeley and the National Facility ELECMI ICTS node ``Laboratorio de Microscopías Avanzadas" at the University of Zaragoza for access to the helium ion microscope and scanning electron microscope facilities, respectively, and for their technical support. 
C.\ Canari-Chumpitaz and C.\ Diaz-Celis of the Bustamante Laboratory at UC Berkeley are thanked for the DNA and tetrasome samples.
TEM characterization at the Molecular Foundry at Lawrence Berkeley National Laboratory was supported by the Office of Science, Office of Basic Energy Sciences, of the U.S. Department of Energy under Contract No.~DE-AC02-05CH11231. 
B.O. acknowledges financial support from the Howard Hughes Medical Institute.
J.M.D.T. acknowledges financial support from grants PID2020-112914RB-I00 and PDC2021-120852-C21 funded by MCIN/AEI/10.13039/501100011033 and the European Union NextGenerationEU/PRTR, and from grant E13$\_$23R funded by Gobierno de Aragón. 

\end{acknowledgement}

%%%%%%%%%%%%%%%%%%%%%%%%%%%%%%%%%%%%%%%%%%%%%%%%%%%%%%%%%%%%%%%%%%%%%
%% The same is true for Supporting Information, which should use the
%% suppinfo environment.
%%%%%%%%%%%%%%%%%%%%%%%%%%%%%%%%%%%%%%%%%%%%%%%%%%%%%%%%%%%%%%%%%%%%%
%\begin{suppinfo}

%This will usually read something like: ``Experimental procedures and
%characterization data for all new compounds. The class will
%automatically add a sentence pointing to the information on-line:

%\end{suppinfo}

%%%%%%%%%%%%%%%%%%%%%%%%%%%%%%%%%%%%%%%%%%%%%%%%%%%%%%%%%%%%%%%%%%%%%
%% The appropriate \bibliography command should be placed here.
%% Notice that the class file automatically sets \bibliographystyle
%% and also names the section correctly.
%%%%%%%%%%%%%%%%%%%%%%%%%%%%%%%%%%%%%%%%%%%%%%%%%%%%%%%%%%%%%%%%%%%%%
\bibliography{main}

\providecommand{\latin}[1]{#1}
\makeatletter
\providecommand{\doi}
  {\begingroup\let\do\@makeother\dospecials
  \catcode`\{=1 \catcode`\}=2 \doi@aux}
\providecommand{\doi@aux}[1]{\endgroup\texttt{#1}}
\makeatother
\providecommand*\mcitethebibliography{\thebibliography}
\csname @ifundefined\endcsname{endmcitethebibliography}
  {\let\endmcitethebibliography\endthebibliography}{}
\begin{mcitethebibliography}{46}
\providecommand*\natexlab[1]{#1}
\providecommand*\mciteSetBstSublistMode[1]{}
\providecommand*\mciteSetBstMaxWidthForm[2]{}
\providecommand*\mciteBstWouldAddEndPuncttrue
  {\def\EndOfBibitem{\unskip.}}
\providecommand*\mciteBstWouldAddEndPunctfalse
  {\let\EndOfBibitem\relax}
\providecommand*\mciteSetBstMidEndSepPunct[3]{}
\providecommand*\mciteSetBstSublistLabelBeginEnd[3]{}
\providecommand*\EndOfBibitem{}
\mciteSetBstSublistMode{f}
\mciteSetBstMaxWidthForm{subitem}{(\alph{mcitesubitemcount})}
\mciteSetBstSublistLabelBeginEnd
  {\mcitemaxwidthsubitemform\space}
  {\relax}
  {\relax}

\bibitem[Ando \latin{et~al.}(2013)Ando, Uchihashi, and Kodera]{Ando2013}
Ando,~T.; Uchihashi,~T.; Kodera,~N. High-speed {AFM} and applications to
  biomolecular systems. \emph{Annu. Rev. Biophys.} \textbf{2013}, \emph{42},
  393--414\relax
\mciteBstWouldAddEndPuncttrue
\mciteSetBstMidEndSepPunct{\mcitedefaultmidpunct}
{\mcitedefaultendpunct}{\mcitedefaultseppunct}\relax
\EndOfBibitem
\bibitem[Ando \latin{et~al.}(2014)Ando, Uchihashi, and Scheuring]{Ando2014}
Ando,~T.; Uchihashi,~T.; Scheuring,~S. Filming biomolecular processes by
  high-speed atomic force microscopy. \emph{Chem. Rev.} \textbf{2014},
  \emph{114}, 3120--3188\relax
\mciteBstWouldAddEndPuncttrue
\mciteSetBstMidEndSepPunct{\mcitedefaultmidpunct}
{\mcitedefaultendpunct}{\mcitedefaultseppunct}\relax
\EndOfBibitem
\bibitem[Fantner \latin{et~al.}(2006)Fantner, Schitter, Kindt, Ivanov, Ivanova,
  Patel, Holten-Andersen, Adams, Thurner, Rangelow, and Hansma]{Fantner2006}
Fantner,~G.~E.; Schitter,~G.; Kindt,~J.~H.; Ivanov,~T.; Ivanova,~K.; Patel,~R.;
  Holten-Andersen,~N.; Adams,~J.; Thurner,~P.~J.; Rangelow,~I.~W.;
  Hansma,~P.~K. Components for high speed atomic force microscopy.
  \emph{Ultramicroscopy} \textbf{2006}, \emph{106}, 881--887\relax
\mciteBstWouldAddEndPuncttrue
\mciteSetBstMidEndSepPunct{\mcitedefaultmidpunct}
{\mcitedefaultendpunct}{\mcitedefaultseppunct}\relax
\EndOfBibitem
\bibitem[Brown \latin{et~al.}(2013)Brown, Picco, Miles, and Faul]{Brown2013}
Brown,~B.~P.; Picco,~L.; Miles,~M.~J.; Faul,~C. F.~J. Opportunities in
  high-speed atomic force microscopy. \emph{Small} \textbf{2013}, \emph{9},
  3201--3211\relax
\mciteBstWouldAddEndPuncttrue
\mciteSetBstMidEndSepPunct{\mcitedefaultmidpunct}
{\mcitedefaultendpunct}{\mcitedefaultseppunct}\relax
\EndOfBibitem
\bibitem[Ando(2018)]{Ando2018}
Ando,~T. High-speed atomic force microscopy and its future prospects.
  \emph{Biophys. Rev.} \textbf{2018}, \emph{10}, 285--292\relax
\mciteBstWouldAddEndPuncttrue
\mciteSetBstMidEndSepPunct{\mcitedefaultmidpunct}
{\mcitedefaultendpunct}{\mcitedefaultseppunct}\relax
\EndOfBibitem
\bibitem[Fukuda and Ando(2021)Fukuda, and Ando]{Fukuda2021}
Fukuda,~S.; Ando,~T. Faster high-speed atomic force microscopy for imaging of
  biomolecular processes. \emph{Rev. Sci. Instrum.} \textbf{2021}, \emph{92},
  033705\relax
\mciteBstWouldAddEndPuncttrue
\mciteSetBstMidEndSepPunct{\mcitedefaultmidpunct}
{\mcitedefaultendpunct}{\mcitedefaultseppunct}\relax
\EndOfBibitem
\bibitem[Ding \latin{et~al.}(2022)Ding, Kuang, Xiong, Mao, Xu, Wang, and
  Hu]{Ding2022}
Ding,~X.; Kuang,~B.; Xiong,~C.; Mao,~R.; Xu,~Y.; Wang,~Z.; Hu,~H. A super high
  aspect ratio atomic force microscopy probe for accurate topography and
  surface tension measurement. \emph{Sens. Actuators A Phys.} \textbf{2022},
  \emph{347}, 113891\relax
\mciteBstWouldAddEndPuncttrue
\mciteSetBstMidEndSepPunct{\mcitedefaultmidpunct}
{\mcitedefaultendpunct}{\mcitedefaultseppunct}\relax
\EndOfBibitem
\bibitem[Ageev \latin{et~al.}(2015)Ageev, Kolomiytsev, Bykov, Smirnov, and
  Kots]{Ageev2015}
Ageev,~O.~A.; Kolomiytsev,~A.~S.; Bykov,~A.~V.; Smirnov,~V.~A.; Kots,~I.~N.
  Fabrication of advanced probes for atomic force microscopy using focused ion
  beam. \emph{Microelectron. Reliab.} \textbf{2015}, \emph{55},
  2131--2134\relax
\mciteBstWouldAddEndPuncttrue
\mciteSetBstMidEndSepPunct{\mcitedefaultmidpunct}
{\mcitedefaultendpunct}{\mcitedefaultseppunct}\relax
\EndOfBibitem
\bibitem[Wilson and Macpherson(2009)Wilson, and Macpherson]{Wilson2009}
Wilson,~N.~R.; Macpherson,~J.~V. Carbon nanotube tips for atomic force
  microscopy. \emph{Nat. Nanotechnol.} \textbf{2009}, \emph{4}, 483--491\relax
\mciteBstWouldAddEndPuncttrue
\mciteSetBstMidEndSepPunct{\mcitedefaultmidpunct}
{\mcitedefaultendpunct}{\mcitedefaultseppunct}\relax
\EndOfBibitem
\bibitem[Engstrom \latin{et~al.}(2011)Engstrom, Savu, Zhu, Bu, Milne, Brugger,
  and Boggild]{Engstrom2011}
Engstrom,~D.~S.; Savu,~V.; Zhu,~X.; Bu,~I. Y.~Y.; Milne,~W.~I.; Brugger,~J.;
  Boggild,~P. High throughput nanofabrication of silicon nanowire and carbon
  nanotube tips on {AFM} probes by stencil-deposited catalysts. \emph{Nano
  Lett.} \textbf{2011}, \emph{11}, 1568--1574\relax
\mciteBstWouldAddEndPuncttrue
\mciteSetBstMidEndSepPunct{\mcitedefaultmidpunct}
{\mcitedefaultendpunct}{\mcitedefaultseppunct}\relax
\EndOfBibitem
\bibitem[Plank \latin{et~al.}(2020)Plank, Winkler, Schwalb, H{\"u}tner,
  Fowlkes, Rack, Utke, and Huth]{Plank2020}
Plank,~H.; Winkler,~R.; Schwalb,~C.~H.; H{\"u}tner,~J.; Fowlkes,~J.~D.;
  Rack,~P.~D.; Utke,~I.; Huth,~M. Focused Electron {Beam-Based} {3D}
  Nanoprinting for Scanning Probe Microscopy: A Review. \emph{Micromachines
  (Basel)} \textbf{2020}, \emph{11}, 48\relax
\mciteBstWouldAddEndPuncttrue
\mciteSetBstMidEndSepPunct{\mcitedefaultmidpunct}
{\mcitedefaultendpunct}{\mcitedefaultseppunct}\relax
\EndOfBibitem
\bibitem[Uchihashi \latin{et~al.}(2012)Uchihashi, Kodera, and
  Ando]{Uchihashi2012}
Uchihashi,~T.; Kodera,~N.; Ando,~T. Guide to video recording of structure
  dynamics and dynamic processes of proteins by high-speed atomic force
  microscopy. \emph{Nat. Protoc.} \textbf{2012}, \emph{7}, 1193--1206\relax
\mciteBstWouldAddEndPuncttrue
\mciteSetBstMidEndSepPunct{\mcitedefaultmidpunct}
{\mcitedefaultendpunct}{\mcitedefaultseppunct}\relax
\EndOfBibitem
\bibitem[Utke \latin{et~al.}(2008)Utke, Hoffmann, and Melngailis]{Utke2008}
Utke,~I.; Hoffmann,~P.; Melngailis,~J. Gas-assisted focused electron beam and
  ion beam processing and fabrication. \emph{J. Vac. Sci. Technol. B
  Microelectron. Nanometer Struct. Process. Meas. Phenom.} \textbf{2008},
  \emph{26}, 1197--1276\relax
\mciteBstWouldAddEndPuncttrue
\mciteSetBstMidEndSepPunct{\mcitedefaultmidpunct}
{\mcitedefaultendpunct}{\mcitedefaultseppunct}\relax
\EndOfBibitem
\bibitem[Akama \latin{et~al.}(1990)Akama, Nishimura, Sakai, and
  Murakami]{Akama1990}
Akama,~Y.; Nishimura,~E.; Sakai,~A.; Murakami,~H. New scanning tunneling
  microscopy tip for measuring surface topography. \emph{J. Vac. Sci. Technol.
  A} \textbf{1990}, \emph{8}, 429--433\relax
\mciteBstWouldAddEndPuncttrue
\mciteSetBstMidEndSepPunct{\mcitedefaultmidpunct}
{\mcitedefaultendpunct}{\mcitedefaultseppunct}\relax
\EndOfBibitem
\bibitem[Walters \latin{et~al.}(1996)Walters, Cleveland, Thomson, Hansma,
  Wendman, Gurley, and Elings]{Walters1996}
Walters,~D.~A.; Cleveland,~J.~P.; Thomson,~N.~H.; Hansma,~P.~K.;
  Wendman,~M.~A.; Gurley,~G.; Elings,~V. Short cantilevers for atomic force
  microscopy. \emph{Rev. Sci. Instrum.} \textbf{1996}, \emph{67},
  3583--3590\relax
\mciteBstWouldAddEndPuncttrue
\mciteSetBstMidEndSepPunct{\mcitedefaultmidpunct}
{\mcitedefaultendpunct}{\mcitedefaultseppunct}\relax
\EndOfBibitem
\bibitem[Orús \latin{et~al.}(2020)Orús, Córdoba, and De~Teresa]{Orus2020}
Orús,~P.; Córdoba,~R.; De~Teresa,~J.~M. \emph{Nanofabrication}; 2053-2563;
  IOP Publishing, 2020; pp 5--1 to 5--58\relax
\mciteBstWouldAddEndPuncttrue
\mciteSetBstMidEndSepPunct{\mcitedefaultmidpunct}
{\mcitedefaultendpunct}{\mcitedefaultseppunct}\relax
\EndOfBibitem
\bibitem[De~Teresa \latin{et~al.}(2009)De~Teresa, C{\'o}rdoba,
  Fern{\'a}ndez-Pacheco, Montero, Strichovanec, and Ibarra]{De_Teresa2009}
De~Teresa,~J.~M.; C{\'o}rdoba,~R.; Fern{\'a}ndez-Pacheco,~A.; Montero,~O.;
  Strichovanec,~P.; Ibarra,~M.~R. Origin of the difference in the resistivity
  of as-grown focused-ion- and focused-electron-beam-induced Pt nanodeposits.
  \emph{J. Nanomater.} \textbf{2009}, \emph{2009}, 1--11\relax
\mciteBstWouldAddEndPuncttrue
\mciteSetBstMidEndSepPunct{\mcitedefaultmidpunct}
{\mcitedefaultendpunct}{\mcitedefaultseppunct}\relax
\EndOfBibitem
\bibitem[Allen(2021)]{Allen2021b}
Allen,~F.~I. A review of defect engineering, ion implantation, and
  nanofabrication using the helium ion microscope. \emph{Beilstein J.
  Nanotechnol.} \textbf{2021}, \emph{12}, 633--664\relax
\mciteBstWouldAddEndPuncttrue
\mciteSetBstMidEndSepPunct{\mcitedefaultmidpunct}
{\mcitedefaultendpunct}{\mcitedefaultseppunct}\relax
\EndOfBibitem
\bibitem[Alkemade and Miro(2014)Alkemade, and Miro]{Alkemade2014}
Alkemade,~P. F.~A.; Miro,~H. Focused helium-ion-beam-induced deposition.
  \emph{Appl. Phys. A: Mater. Sci. Process.} \textbf{2014}, \emph{117},
  1727--1747\relax
\mciteBstWouldAddEndPuncttrue
\mciteSetBstMidEndSepPunct{\mcitedefaultmidpunct}
{\mcitedefaultendpunct}{\mcitedefaultseppunct}\relax
\EndOfBibitem
\bibitem[C{\'o}rdoba \latin{et~al.}(2018)C{\'o}rdoba, Ibarra, Mailly, and
  De~Teresa]{Cordoba2018}
C{\'o}rdoba,~R.; Ibarra,~A.; Mailly,~D.; De~Teresa,~J.~M. Vertical Growth of
  Superconducting Crystalline Hollow Nanowires by He+ Focused Ion Beam Induced
  Deposition. \emph{Nano Lett.} \textbf{2018}, \emph{18}, 1379--1386\relax
\mciteBstWouldAddEndPuncttrue
\mciteSetBstMidEndSepPunct{\mcitedefaultmidpunct}
{\mcitedefaultendpunct}{\mcitedefaultseppunct}\relax
\EndOfBibitem
\bibitem[C{\'o}rdoba \latin{et~al.}(2019)C{\'o}rdoba, Mailly, Rezaev, Smirnova,
  Schmidt, Fomin, Zeitler, Guillam{\'o}n, Suderow, and De~Teresa]{Cordoba2019}
C{\'o}rdoba,~R.; Mailly,~D.; Rezaev,~R.~O.; Smirnova,~E.~I.; Schmidt,~O.~G.;
  Fomin,~V.~M.; Zeitler,~U.; Guillam{\'o}n,~I.; Suderow,~H.; De~Teresa,~J.~M.
  {Three-Dimensional} Superconducting Nanohelices Grown by
  {He+-Focused-Ion-Beam} Direct Writing. \emph{Nano Lett.} \textbf{2019},
  \emph{19}, 8597--8604\relax
\mciteBstWouldAddEndPuncttrue
\mciteSetBstMidEndSepPunct{\mcitedefaultmidpunct}
{\mcitedefaultendpunct}{\mcitedefaultseppunct}\relax
\EndOfBibitem
\bibitem[Nanda \latin{et~al.}(2015)Nanda, van Veldhoven, Maas, Sadeghian, and
  Alkemade]{Nanda2015}
Nanda,~G.; van Veldhoven,~E.; Maas,~D.; Sadeghian,~H.; Alkemade,~P. F.~A.
  Helium ion beam induced growth of hammerhead {AFM} probes. \emph{J. Vac. Sci.
  Technol. B Microelectron. Nanometer Struct. Process. Meas. Phenom.}
  \textbf{2015}, \emph{33}, 06F503\relax
\mciteBstWouldAddEndPuncttrue
\mciteSetBstMidEndSepPunct{\mcitedefaultmidpunct}
{\mcitedefaultendpunct}{\mcitedefaultseppunct}\relax
\EndOfBibitem
\bibitem[Jaafar \latin{et~al.}(2020)Jaafar, Pablo-Navarro, Berganza, Ares,
  Mag{\'e}n, Masseboeuf, Gatel, Snoeck, G{\'o}mez-Herrero, de~Teresa, and
  Asenjo]{Jaafar2020}
Jaafar,~M.; Pablo-Navarro,~J.; Berganza,~E.; Ares,~P.; Mag{\'e}n,~C.;
  Masseboeuf,~A.; Gatel,~C.; Snoeck,~E.; G{\'o}mez-Herrero,~J.;
  de~Teresa,~J.~M.; Asenjo,~A. Customized {MFM} probes based on magnetic
  nanorods. \emph{Nanoscale} \textbf{2020}, \emph{12}, 10090--10097\relax
\mciteBstWouldAddEndPuncttrue
\mciteSetBstMidEndSepPunct{\mcitedefaultmidpunct}
{\mcitedefaultendpunct}{\mcitedefaultseppunct}\relax
\EndOfBibitem
\bibitem[Winkler \latin{et~al.}(2023)Winkler, Ciria, Ahmad, Plank, and
  Marcuello]{Winkler2023}
Winkler,~R.; Ciria,~M.; Ahmad,~M.; Plank,~H.; Marcuello,~C. A Review of the
  Current State of Magnetic Force Microscopy to Unravel the Magnetic Properties
  of Nanomaterials Applied in Biological Systems and Future Directions for
  Quantum Technologies. \emph{Nanomaterials (Basel)} \textbf{2023},
  \emph{13}\relax
\mciteBstWouldAddEndPuncttrue
\mciteSetBstMidEndSepPunct{\mcitedefaultmidpunct}
{\mcitedefaultendpunct}{\mcitedefaultseppunct}\relax
\EndOfBibitem
\bibitem[Allen(2021)]{Allen2021}
Allen,~F.~I. Branched High Aspect Ratio Nanostructures Fabricated by Focused
  Helium Ion Beam Induced Deposition of an Insulator. \emph{Micromachines
  (Basel)} \textbf{2021}, \emph{12}\relax
\mciteBstWouldAddEndPuncttrue
\mciteSetBstMidEndSepPunct{\mcitedefaultmidpunct}
{\mcitedefaultendpunct}{\mcitedefaultseppunct}\relax
\EndOfBibitem
\bibitem[Allen \latin{et~al.}(2020)Allen, Hosemann, and Balooch]{Allen2020}
Allen,~F.~I.; Hosemann,~P.; Balooch,~M. Key mechanistic features of swelling
  and blistering of helium-ion-irradiated tungsten. \emph{Scr. Mater.}
  \textbf{2020}, \emph{178}, 256--260\relax
\mciteBstWouldAddEndPuncttrue
\mciteSetBstMidEndSepPunct{\mcitedefaultmidpunct}
{\mcitedefaultendpunct}{\mcitedefaultseppunct}\relax
\EndOfBibitem
\bibitem[Hlawacek \latin{et~al.}(2013)Hlawacek, Ahmad, Smithers, and
  Kooij]{Hlawacek2013}
Hlawacek,~G.; Ahmad,~I.; Smithers,~M.~A.; Kooij,~E.~S. To see or not to see:
  imaging surfactant coated nano-particles using {HIM} and {SEM}.
  \emph{Ultramicroscopy} \textbf{2013}, \emph{135}, 89--94\relax
\mciteBstWouldAddEndPuncttrue
\mciteSetBstMidEndSepPunct{\mcitedefaultmidpunct}
{\mcitedefaultendpunct}{\mcitedefaultseppunct}\relax
\EndOfBibitem
\bibitem[Barr and Seal(1995)Barr, and Seal]{Barr1995}
Barr,~T.~L.; Seal,~S. Nature of the use of adventitious carbon as a binding
  energy standard. \emph{J. Vac. Sci. Technol. A} \textbf{1995}, \emph{13},
  1239--1246\relax
\mciteBstWouldAddEndPuncttrue
\mciteSetBstMidEndSepPunct{\mcitedefaultmidpunct}
{\mcitedefaultendpunct}{\mcitedefaultseppunct}\relax
\EndOfBibitem
\bibitem[Utke \latin{et~al.}(2020)Utke, Michler, Winkler, and Plank]{Utke2020}
Utke,~I.; Michler,~J.; Winkler,~R.; Plank,~H. Mechanical Properties of {3D}
  Nanostructures Obtained by Focused {Electron/Ion} {Beam-Induced} Deposition:
  A Review. \emph{Micromachines (Basel)} \textbf{2020}, \emph{11}\relax
\mciteBstWouldAddEndPuncttrue
\mciteSetBstMidEndSepPunct{\mcitedefaultmidpunct}
{\mcitedefaultendpunct}{\mcitedefaultseppunct}\relax
\EndOfBibitem
\bibitem[Strigul \latin{et~al.}(2005)Strigul, Koutsospyros, Arienti,
  Christodoulatos, Dermatas, and Braida]{Strigul2005}
Strigul,~N.; Koutsospyros,~A.; Arienti,~P.; Christodoulatos,~C.; Dermatas,~D.;
  Braida,~W. Effects of tungsten on environmental systems. \emph{Chemosphere}
  \textbf{2005}, \emph{61}, 248--258\relax
\mciteBstWouldAddEndPuncttrue
\mciteSetBstMidEndSepPunct{\mcitedefaultmidpunct}
{\mcitedefaultendpunct}{\mcitedefaultseppunct}\relax
\EndOfBibitem
\bibitem[Seiffert \latin{et~al.}(2007)Seiffert, Ullmann, Messerschmidt, Schink,
  Kroneck, and Einsle]{Seiffert2007}
Seiffert,~G.~B.; Ullmann,~G.~M.; Messerschmidt,~A.; Schink,~B.; Kroneck,~P.
  M.~H.; Einsle,~O. Structure of the non-redox-active {tungsten/[4Fe:4S}]
  enzyme acetylene hydratase. \emph{Proc. Natl. Acad. Sci. U. S. A.}
  \textbf{2007}, \emph{104}, 3073--3077\relax
\mciteBstWouldAddEndPuncttrue
\mciteSetBstMidEndSepPunct{\mcitedefaultmidpunct}
{\mcitedefaultendpunct}{\mcitedefaultseppunct}\relax
\EndOfBibitem
\bibitem[Kanellis and Dos~Remedios(2018)Kanellis, and
  Dos~Remedios]{Kanellis2018}
Kanellis,~V.~G.; Dos~Remedios,~C.~G. A review of heavy metal cation binding to
  deoxyribonucleic acids for the creation of chemical sensors. \emph{Biophys.
  Rev.} \textbf{2018}, \emph{10}, 1401--1414\relax
\mciteBstWouldAddEndPuncttrue
\mciteSetBstMidEndSepPunct{\mcitedefaultmidpunct}
{\mcitedefaultendpunct}{\mcitedefaultseppunct}\relax
\EndOfBibitem
\bibitem[Drew \latin{et~al.}(1981)Drew, Wing, Takano, Broka, Tanaka, Itakura,
  and Dickerson]{Drew1981}
Drew,~H.~R.; Wing,~R.~M.; Takano,~T.; Broka,~C.; Tanaka,~S.; Itakura,~K.;
  Dickerson,~R.~E. Structure of a {B-DNA} dodecamer: conformation and dynamics.
  \emph{Proc. Natl. Acad. Sci. U. S. A.} \textbf{1981}, \emph{78},
  2179--2183\relax
\mciteBstWouldAddEndPuncttrue
\mciteSetBstMidEndSepPunct{\mcitedefaultmidpunct}
{\mcitedefaultendpunct}{\mcitedefaultseppunct}\relax
\EndOfBibitem
\bibitem[Mart{\'\i}nez \latin{et~al.}(2011)Mart{\'\i}nez, Tello, D{\'\i}az,
  Rom{\'a}n, Garcia, and Huttel]{Martinez2011}
Mart{\'\i}nez,~L.; Tello,~M.; D{\'\i}az,~M.; Rom{\'a}n,~E.; Garcia,~R.;
  Huttel,~Y. Aspect-ratio and lateral-resolution enhancement in force
  microscopy by attaching nanoclusters generated by an ion cluster source at
  the end of a silicon tip. \emph{Rev. Sci. Instrum.} \textbf{2011},
  \emph{82}\relax
\mciteBstWouldAddEndPuncttrue
\mciteSetBstMidEndSepPunct{\mcitedefaultmidpunct}
{\mcitedefaultendpunct}{\mcitedefaultseppunct}\relax
\EndOfBibitem
\bibitem[Lyubchenko(2012)]{Lyubchenko2012}
Lyubchenko,~Y.~L. \emph{{AFM} imaging in liquid of {DNA} and {protein--DNA}
  complexes}; Wiley, 2012; Chapter 9, pp 231--258\relax
\mciteBstWouldAddEndPuncttrue
\mciteSetBstMidEndSepPunct{\mcitedefaultmidpunct}
{\mcitedefaultendpunct}{\mcitedefaultseppunct}\relax
\EndOfBibitem
\bibitem[Prokhorov \latin{et~al.}(2021)Prokhorov, Barinov, Prusakov, Dubrovin,
  Frank-Kamenetskii, and Klinov]{Prokhorov2021}
Prokhorov,~V.~V.; Barinov,~N.~A.; Prusakov,~K.~A.; Dubrovin,~E.~V.;
  Frank-Kamenetskii,~M.~D.; Klinov,~D.~V. Anomalous Laterally Stressed
  Kinetically Trapped {DNA} Surface Conformations. \emph{Nanomicro Lett}
  \textbf{2021}, \emph{13}, 130\relax
\mciteBstWouldAddEndPuncttrue
\mciteSetBstMidEndSepPunct{\mcitedefaultmidpunct}
{\mcitedefaultendpunct}{\mcitedefaultseppunct}\relax
\EndOfBibitem
\bibitem[Japaridze \latin{et~al.}(2016)Japaridze, Vobornik, Lipiec, Cerreta,
  Szczerbinski, Zenobi, and Dietler]{Japaridze2016}
Japaridze,~A.; Vobornik,~D.; Lipiec,~E.; Cerreta,~A.; Szczerbinski,~J.;
  Zenobi,~R.; Dietler,~G. Toward an Effective Control of {DNA's} Submolecular
  Conformation on a Surface. \emph{Macromolecules} \textbf{2016}, \emph{49},
  643--652\relax
\mciteBstWouldAddEndPuncttrue
\mciteSetBstMidEndSepPunct{\mcitedefaultmidpunct}
{\mcitedefaultendpunct}{\mcitedefaultseppunct}\relax
\EndOfBibitem
\bibitem[Heenan and Perkins(2019)Heenan, and Perkins]{Heenan2019}
Heenan,~P.~R.; Perkins,~T.~T. Imaging {DNA} Equilibrated onto Mica in Liquid
  Using Biochemically Relevant Deposition Conditions. \emph{ACS Nano}
  \textbf{2019}, \emph{13}, 4220--4229\relax
\mciteBstWouldAddEndPuncttrue
\mciteSetBstMidEndSepPunct{\mcitedefaultmidpunct}
{\mcitedefaultendpunct}{\mcitedefaultseppunct}\relax
\EndOfBibitem
\bibitem[Kohli(2019)]{Kohli2019}
Kohli,~R. In \emph{Developments in Surface Contamination and Cleaning:
  Applications of Cleaning Techniques}; Kohli,~R., Mittal,~K.~L., Eds.;
  Elsevier, 2019; pp 355--390\relax
\mciteBstWouldAddEndPuncttrue
\mciteSetBstMidEndSepPunct{\mcitedefaultmidpunct}
{\mcitedefaultendpunct}{\mcitedefaultseppunct}\relax
\EndOfBibitem
\bibitem[Seewald \latin{et~al.}(2022)Seewald, Sattelkow, Brugger-Hatzl,
  Kothleitner, Frerichs, Schwalb, Hummel, and Plank]{Seewald2022}
Seewald,~L.~M.; Sattelkow,~J.; Brugger-Hatzl,~M.; Kothleitner,~G.;
  Frerichs,~H.; Schwalb,~C.; Hummel,~S.; Plank,~H. {3D} Nanoprinting of
  {All-Metal} Nanoprobes for Electric {AFM} Modes. \emph{Nanomaterials (Basel)}
  \textbf{2022}, \emph{12}\relax
\mciteBstWouldAddEndPuncttrue
\mciteSetBstMidEndSepPunct{\mcitedefaultmidpunct}
{\mcitedefaultendpunct}{\mcitedefaultseppunct}\relax
\EndOfBibitem
\bibitem[Epelle \latin{et~al.}(2023)Epelle, Macfarlane, Cusack, Burns, Okolie,
  Mackay, Rateb, and Yaseen]{Epelle2023}
Epelle,~E.~I.; Macfarlane,~A.; Cusack,~M.; Burns,~A.; Okolie,~J.~A.;
  Mackay,~W.; Rateb,~M.; Yaseen,~M. Ozone application in different industries:
  A review of recent developments. \emph{Chem. Eng. J.} \textbf{2023},
  \emph{454}, 140188\relax
\mciteBstWouldAddEndPuncttrue
\mciteSetBstMidEndSepPunct{\mcitedefaultmidpunct}
{\mcitedefaultendpunct}{\mcitedefaultseppunct}\relax
\EndOfBibitem
\bibitem[Allen \latin{et~al.}(2023)Allen, Blanchard, Lake, Pappas, Xia, Notte,
  Zhang, Minor, and Sanford]{Allen2023}
Allen,~F.~I.; Blanchard,~P.~T.; Lake,~R.; Pappas,~D.; Xia,~D.; Notte,~J.~A.;
  Zhang,~R.; Minor,~A.~M.; Sanford,~N.~A. Fabrication of Specimens for Atom
  Probe Tomography Using a Combined Gallium and Neon Focused Ion Beam Milling
  Approach. \emph{Microsc. Microanal.} \textbf{2023}, \emph{29},
  1628--1638\relax
\mciteBstWouldAddEndPuncttrue
\mciteSetBstMidEndSepPunct{\mcitedefaultmidpunct}
{\mcitedefaultendpunct}{\mcitedefaultseppunct}\relax
\EndOfBibitem
\bibitem[Wendel \latin{et~al.}(1995)Wendel, Lorenz, and Kotthaus]{Wendel1995}
Wendel,~M.; Lorenz,~H.; Kotthaus,~J.~P. Sharpened electron beam deposited tips
  for high resolution atomic force microscope lithography and imaging.
  \emph{Appl. Phys. Lett.} \textbf{1995}, \emph{67}, 3732--3734\relax
\mciteBstWouldAddEndPuncttrue
\mciteSetBstMidEndSepPunct{\mcitedefaultmidpunct}
{\mcitedefaultendpunct}{\mcitedefaultseppunct}\relax
\EndOfBibitem
\bibitem[Kometani and Katsuda(2021)Kometani, and Katsuda]{Kometani2021}
Kometani,~R.; Katsuda,~M. Miniaturization process of three-dimensional
  nanostructures fabricated by focused-ion-beam chemical vapor deposition.
  \emph{Jpn. J. Appl. Phys.} \textbf{2021}, \emph{60}, 128002\relax
\mciteBstWouldAddEndPuncttrue
\mciteSetBstMidEndSepPunct{\mcitedefaultmidpunct}
{\mcitedefaultendpunct}{\mcitedefaultseppunct}\relax
\EndOfBibitem
\bibitem[Hill and Faridur~Rahman(2011)Hill, and Faridur~Rahman]{Hill2011}
Hill,~R.; Faridur~Rahman,~F. H.~M. Advances in helium ion microscopy.
  \emph{Nucl. Instrum. Methods Phys. Res. A} \textbf{2011}, \emph{645},
  96--101\relax
\mciteBstWouldAddEndPuncttrue
\mciteSetBstMidEndSepPunct{\mcitedefaultmidpunct}
{\mcitedefaultendpunct}{\mcitedefaultseppunct}\relax
\EndOfBibitem
\end{mcitethebibliography}

\end{document}

\newpage

\begin{figure*}[]
  \caption{Dark-field STEM (left) and STEM-XEDS elemental maps (right) for He-FIBID tip showing distribution of W, C and O (note color scheme is different compared with main manuscript). The nanopillar width in the C and O maps is larger than in the W map, indicating a surface layer composed of C and O (thickness \unit[$\sim$3]{nm}). Z-contrast in dark-field STEM means that most of the contrast in that image comes from W, explaining why the nanopillar widths in the dark-field STEM and W maps are essentially the same.}
  \includegraphics[width=0.4\linewidth]{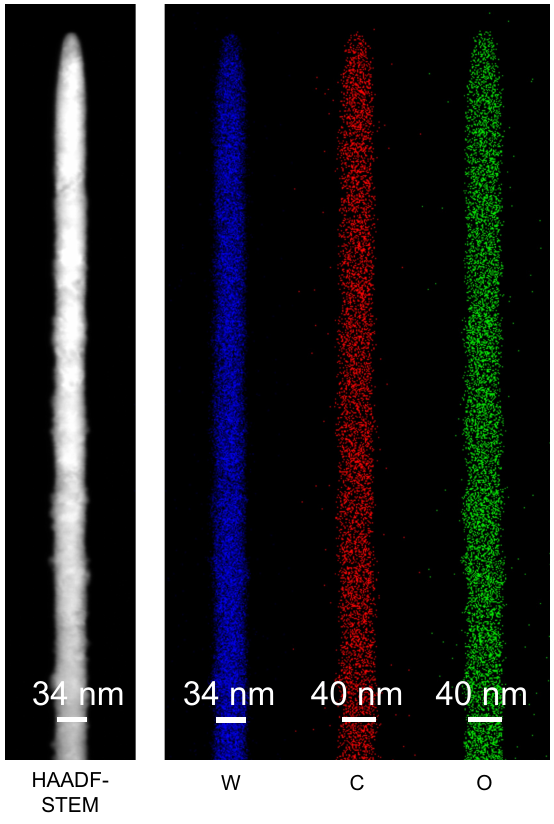}
  \label{fgr:S1}
\end{figure*}

\begin{figure*}[]
  \caption{He-FIB milling strategy to thin He-FIBID nanopillar from opposite sides. The blue rectangles denote the milling patterns used. Imaging by HIM.}
  \includegraphics[width=0.5\linewidth]{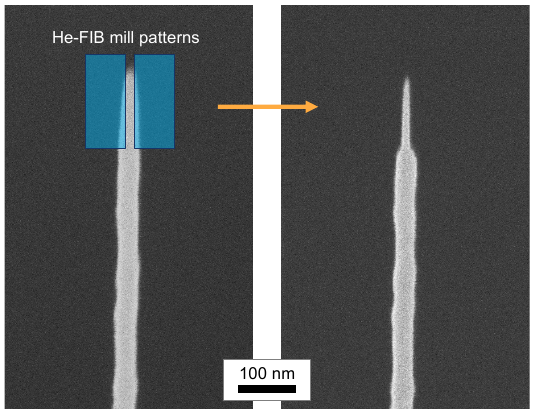}
  \label{fgr:S2}
\end{figure*}

\begin{table*}
\begin{tabular}{ |p{3cm}|p{3cm}|p{3cm}|p{3cm}|  }
 \hline
 Tip used & Point on DNA & Height [nm] & FWHM [nm]\\
 \hline
 He-FIBID  & 1 & 1.5 & 8.7\\
 & 2 & 1.8 & 9.8\\
 \hline
 Commercial & 1 & 1.2 & 7.0\\
 & 2 & 1.2 & 8.2\\
 \hline
\end{tabular}
\caption{Line profile analysis for Fig.~3, investigating an organometallic He-FIBID tip vs.\ a commercial high-density-carbon tip scanning DNA fragment at \unit[1]{f/s}. The commercial tip was a cone-shaped tip of a new Type I cantilever, whereas the He-FIBID tip was grown onto the blunted cone of a previously-used Type I cantilever. The tip performances are comparable.}
\end{table*}

\begin{table*}
\begin{tabular}{ |p{2cm}|p{1.5cm}|p{1.5cm}|p{2cm}|p{2cm}|p{2cm}|p{2cm}|}
 \hline
 DNA region scanned & Scan \# & Scan rate [f/s] & \multicolumn{2}{|l|}{Height [nm]} & \multicolumn{2}{|l|}{FWHM [nm]}\\\cline{4-7}
                     &    &   & Point 1   &  Point 2  &  Point 1  &  Point 2 \\
 \hline     
 \multirow{2}{*}{Full length} & first  & \multirow{2}{*}{1} & 1.7 & 1.1 & 6.1 & 7.6\\
 & last   &   & 1.1 & 0.9 & 9.3 & 6.7\\
 \hline
 \multirow{2}{*}{Loop} & first         & \multirow{2}{*}{2} & 1.9 & 1.6 & 6.5 & 6.5\\
 & last           &   & 1.7 & 1.8 & 6.2 & 5.5\\
 \hline
 \multirow{2}{*}{Loop} & first          & \multirow{2}{*}{5} & 2.2 & 1.9 & 6.0 & 5.1\\
 & last           &   & 2.1 & 1.8 & 5.8 & 4.4\\
 \hline
 \multirow{2}{*}{Loop} & first         & \multirow{2}{*}{10} & 2.0 & 1.6 & 4.9 & 5.1\\
 & last          &    & 1.8 & 1.8 & 6.1 & 5.0\\
 \hline
\end{tabular}
\caption{Line profile analysis for Fig.~4, investigating an organometallic FEBID tip grown on a Type II cantilever repeatedly scanning a long length of DNA at different frame rates and for different fields of view (same tip used throughout).}
\end{table*}

\begin{table*}
\begin{tabular}{ |p{6cm}|p{3cm}|p{3cm}|p{3cm}|  }
 \hline
 Experiment & Point on DNA & Height [nm] & FWHM [nm]\\
 \hline
 \multirow{2}{*}{1$^{st}$ tetrasome sample} & 1 & 0.95 & 6.6\\
 & 2 & 0.88 & 5.2\\
 \hline
 \multirow{2}{*}{2$^{nd}$ tetrasome sample} & 1 & 0.73 & 8.9\\
 & 2 & 0.76 & 8.7\\
 \hline
\end{tabular}
\caption{Line profile analysis for Figs.~6b and 6d, investigating the same organometallic He-FIBID tip scanning different tetrasome samples, having imaged the tip by HIM before and after each HS-AFM experiment.}
\end{table*}

%%%%%%%%%%%%%%%%%%%%%%%%%%%%%%%%%%%%%%%%%%%%%%%%%%%%%%%%%%%%%%%%%%%%%
%% The appropriate \bibliography command should be placed here.
%% Notice that the class file automatically sets \bibliographystyle
%% and also names the section correctly.
%%%%%%%%%%%%%%%%%%%%%%%%%%%%%%%%%%%%%%%%%%%%%%%%%%%%%%%%%%%%%%%%%%%%%
%\bibliography{achemso-demo,bib}